\documentclass[11pt]{article}
\usepackage{acl}
\usepackage{times}
\usepackage{latexsym}
\usepackage{amsmath}
\usepackage{amssymb}
\usepackage[T1]{fontenc}
\usepackage[utf8]{inputenc}
\usepackage{microtype}
\usepackage{inconsolata}
\usepackage{graphicx}
\usepackage{booktabs}
\usepackage{multirow}
\usepackage{xcolor}
\usepackage{listings}
\usepackage{url}
\usepackage[noabbrev]{cleveref}

\lstdefinestyle{prompt}{
  basicstyle=\ttfamily\footnotesize,
  frame=single,
  framerule=0.4pt,
  rulecolor=\color{black},
  breaklines=true,
  breakatwhitespace=false,
  breakautoindent=false,
  breakindent=0pt,
  columns=fullflexible,
  keepspaces=true,
  showstringspaces=false,
  upquote=true,
  xleftmargin=0pt,
  xrightmargin=0pt,
  framexleftmargin=3pt,
  framexrightmargin=3pt,
  framextopmargin=2pt,
  framexbottommargin=2pt,
  aboveskip=0.25\baselineskip,
  belowskip=0.25\baselineskip,
  captionpos=b
}

\newcommand{\method}{IGPO}

\title{Inventory-Grounded Policy-Level Optimization for Training-Free \\ AI Search}

\author{
  Wei Zhou\thanks{\ \ Equal contribution to this research (co-first authors).} \quad Tiandeng Wu\footnotemark[1] \quad Jiandong Ding\thanks{\ \ Corresponding author.} \quad Zhufeng Fan \quad Yi Cao \\
  Huawei Technologies Co., Ltd., China \\
  \texttt{\{zhouwei281,wutiandeng1,dingjiandong2\}@huawei.com} \\
  \texttt{\{fanzhufeng,caoyi23\}@huawei.com}
}

\begin{document}
\maketitle

\begin{abstract}
Early in deployment, an AI search system typically operates over a frequently updated product catalog, so the available items and their properties cannot be treated as stable knowledge that can be encoded in fixed prompts or strategies.
Fine-tuning, reinforcement learning, and static prompt patches fit poorly: labels are scarce, rewards drift with inventory, model releases are costly, and prompt fixes quickly stale. We present \textbf{I}nventory-\textbf{G}rounded \textbf{P}olicy-Level \textbf{O}ptimization (\method{}), a training-free approach for fixed AI search pipelines. \method{} separates policy from environment facts: it learns Policy Guidelines for acting on runtime inventory evidence rather than memorizing available items. Online, IGPO grounds each query by probing the inventory and constructing an inventory portrait, then injects relevant Policy Guidelines into the retrieval and selection prompts. Offline, stochastic rollouts are grouped by query -- mixed outcome groups directly yield contrastive signal, and an inventory-guided exploration loop distinguishes missed retrieval routes from cases where no matching support is found under the observed inventory evidence. Since May 2026, \method{} has been deployed in a commercial smart-assistant AI search system.
A 14-day online A/B test of the complete \method{} treatment shows a 3.17\% relative CTR lift and a 38.9\% reduction in audited bad cases.
\end{abstract}

\section{Introduction}

Production AI search systems increasingly map natural-language requests to concrete items from a dynamic, business-constrained inventory. Unlike mature web or e-commerce search, early-stage AI search deployments often operate over rapidly changing catalogs. Items are added, removed, or renamed, and metadata reliability varies across categories. As a result, the catalog cannot serve as a static knowledge source. Its contents and metadata change too often to be encoded into fixed prompts or strategies.

Suppose a user asks “order a birthday cake with photo print” but receives no satisfactory results. The failure could stem from (i) the needed item truly being absent from the inventory, (ii) the retrieval route failing to discover an actually existing matching item, or (iii) the item being retrieved but then not selected by the final ranking/selection module. 
Conventional optimization strategies fit poorly in early-stage AI search systems. Fine-tuning requires labeled data and frequent model releases, which is too heavy for rapid inventory adaptation. Reinforcement learning struggles to define a stable reward when item availability keeps changing. Online reflection or self-correction adds latency and operational risk. Static prompt patches are easier to ship, but they are broad, quickly go stale, and can affect queries far beyond the failure scene. Training-free optimization over execution traces is attractive, but trace-only optimization has a critical blind spot: it may encode transient inventory facts into the policy instead of teaching the policy how to reason from current inventory evidence.

Our premise is that a search policy should not memorize which items currently exist, but learn to reason from the runtime inventory state. We capture this learned decision behavior in a Policy Guideline — a compact, human-readable natural-language artifact that specifies how a decision point should use inventory evidence, retrieval observations, and scene context. A guideline can encode rules such as “if the query mentions photo customization, check the image-printing service catalog, not just physical goods”, or “do not assume express delivery is always available and verify delivery options per item”. Critically, such a guideline does not claim that any particular item is permanently available, but rather separates durable decision procedures from volatile environment facts.

We propose Inventory-Grounded Policy-Level Optimization (IGPO), a training-free framework that improves online performance under dynamic inventory without modifying model weights. Online, IGPO grounds each query by probing the inventory for matching items and constructing a lightweight inventory portrait, then injects relevant Policy Guidelines from a store into the retrieval and selection prompts. Offline, stochastic rollouts are grouped by query. Mixed outcome groups directly yield contrastive improvement signal. For groups that always fail, an inventory-guided exploration loop systematically explores alternative retrieval paths to distinguish cases where a feasible route was initially missed from cases where no matching support is found under the observed inventory evidence. 

Our contributions are:
\begin{itemize}
    \item We formulate post-deployment optimization for early-stage AI search systems with dynamic inventories, distinguishing repairable search failures from cases where no matching support is found under the observed inventory evidence, while decoupling policy from environment facts.
    \item We propose \method{}, a training-free framework that conditions the search pipeline online with inventory evidence and Policy Guidelines and refines those guidelines offline through inventory-guided exploration, without model retraining.
    \item We validate \method{} with offline replay, ablations, a 14-day online A/B test, and sustained deployment in a commercial smart-assistant AI search system.
\end{itemize}

\section{Related Work}
\label{sec:related-work}

\begin{figure*}[t]
\centering
\includegraphics[width=\textwidth]{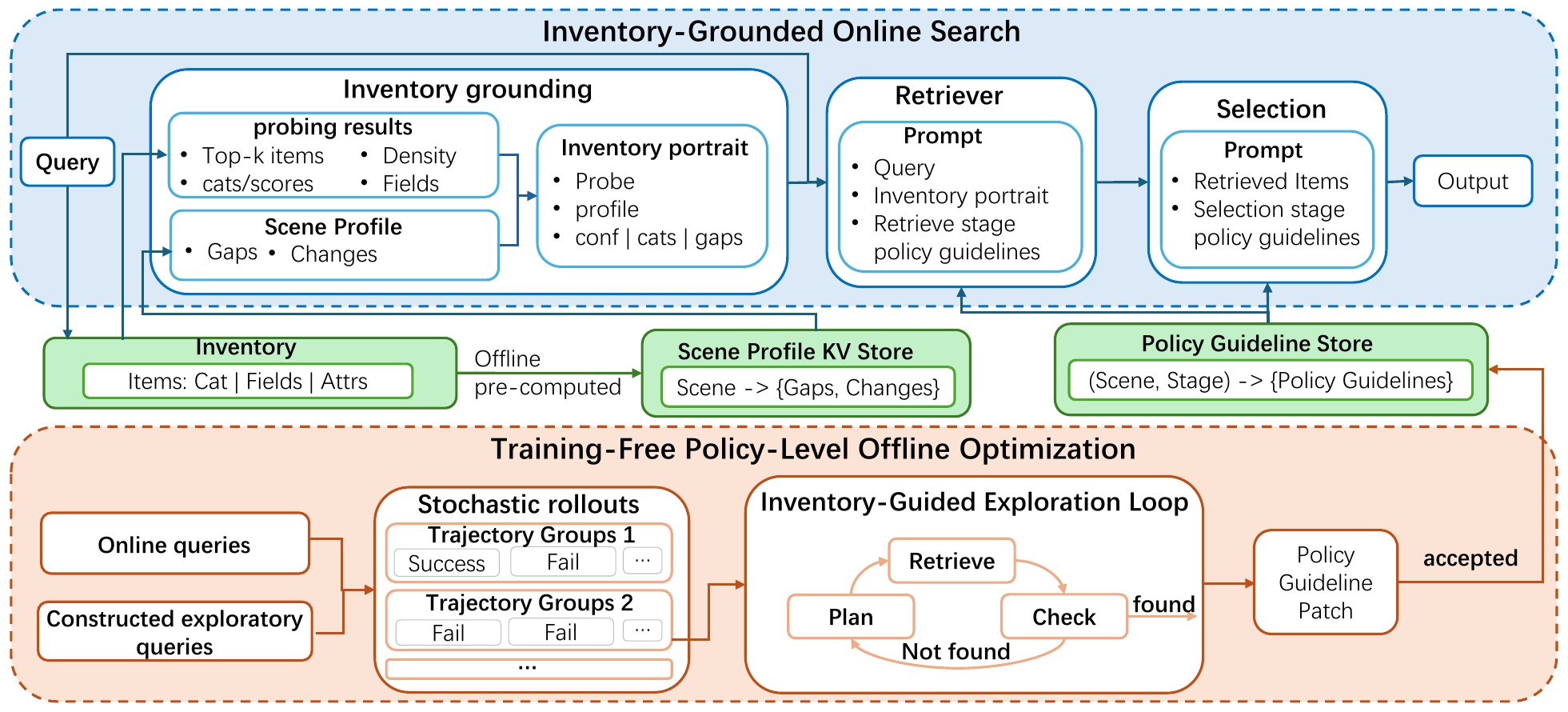}
\caption{Overview of IGPO that online conditions the search pipeline with inventory evidence and Policy Guidelines, and offline refines those guidelines through inventory-guided exploration.
}
\label{fig:method-overview}
\end{figure*}

\paragraph{LLM-based search and query rewriting.}
LLM-based search systems increasingly rely on dense retrieval, retrieval-augmented generation, and agentic reasoning, but they do not directly address post-deployment search-policy optimization when the item inventory itself changes -- items are added, removed, renamed, or regrouped, and metadata reliability varies across categories \citep{karpukhin2020dpr,lewis2020rag,asai2024selfrag,yao2023react,li2025searcho1}. While classic IR research on dynamic collections studies evolving corpora \citep{elsas2010temporal}, and recent query rewriting methods address inventory-aware adaptation in e-commerce search \citep{nguyen2025minielm}, these approaches focus on retrieval effectiveness or lexical alignment rather than post-deployment policy optimization under inventory uncertainty. Weight-updating approaches (e.g., SFT with GRPO \citep{shao2024deepseekmath}) are similarly unsuitable: they require frequent model releases and stable rewards, both undermined by a changing inventory. \method{} instead targets a fixed production pipeline where the central diagnostic challenge is distinguishing search-behavior failures from cases where no matching support is found under the observed inventory evidence.

\paragraph{Training-free prompt and agent optimization.}
Training-free optimization methods improve LLMs by revising prompts or compiled programs using execution traces \citep{yang2024opro,khattab2024dspy,agrawal2026gepa,shinn2023reflexion}. In a dynamic-inventory setting, however, trace-only optimization often encodes transient inventory failures into the revised prompt (e.g., “this category never contains X”) instead of teaching runtime verification, producing a brittle prompt overfitted to a catalog snapshot. \method{} instead learns scene- and stage-indexed Policy Guidelines: conditional rules that consume runtime inventory evidence but never assert the existence of a specific item. This structurally decouples durable policy updates from volatile inventory facts.

\paragraph{Inventory-grounded production search.}
Product search has long aligned queries with catalog facts \citep{lai2018productqa,bagheri2022productrelevance}, and classic IR techniques such as MMR and xQuAD \citep{carbonell1998mmr,santos2010xquad} use retrieved-set information for diversity control, an early form of inventory awareness. \method{} makes inventory evidence a first-class signal in policy optimization: online serving uses inventory portraits synthesized from retrieval probes and scene profiles, while offline updates replay the pipeline against fixed inventory snapshots. This enables failure triage under inventory uncertainty: an inventory-guided exploration loop distinguishes missed-route failures from cases where no matching support is found under the observed inventory evidence, and only validated findings update Policy Guidelines, reducing the risk that unsupported absence conclusions enter the learned policy.

\section{Method}
\label{sec:method}

We aim to improve a fixed LLM-based AI search pipeline under dynamic inventory without modifying model weights or the retriever. The only optimized state is a \textbf{Policy Guideline} store $\mathcal{G}$. Each guideline $g\in\mathcal{G}$ specifies conditional decision behavior for the Retrieval Planner or Selection stage, consuming runtime inventory evidence but never asserting the existence of specific items.

\subsection{System Overview}

IGPO has an online serving path and an offline optimization loop. Online, the frozen Retrieval Planner and the Selection stage are augmented with runtime inventory evidence from a lightweight \emph{inventory portrait} $P(q,\mathcal{I}_t)$ and with stage-relevant Policy Guidelines $\mathcal{G}_{\text{ret}},\mathcal{G}_{\text{sel}}$. The Retrieval Planner issues retrieval routes or takes the \emph{no-inventory path} when evidence indicates no support. Selection then picks final results or produces the configured no-inventory response. Offline, the pipeline is replayed under fixed snapshots to sample trajectories, probe missed routes, and derive validated guideline updates (\cref{sec:offline}), separating policy from transient inventory facts.

Formally, let $\mathcal{Q}$ be the query space and $\mathcal{I}_t$ the inventory at time $t$. The pipeline is:
\begin{equation}\label{eq:pipeline}
\begin{aligned}
\mathcal{C} &= f_{\text{ret}}\big(q,\; P(q, \mathcal{I}_t),\; \mathcal{G}_{\text{ret}}(q, \mathcal{I}_t)\big), \\
\mathcal{R} &= f_{\text{sel}}\big(q,\; \mathcal{C},\; P(q, \mathcal{I}_t),\; \mathcal{G}_{\text{sel}}(q, \mathcal{I}_t)\big),
\end{aligned}
\end{equation}
where $q\in\mathcal{Q}$ is a user query, $f_{\text{ret}},f_{\text{sel}}$ are frozen, $P$ is the inventory portrait, and $\mathcal{G}_{\text{ret}}$ and $\mathcal{G}_{\text{sel}}$ are the query-selected Retrieval- and Selection-stage subsets of $\mathcal{G}$, respectively. The objective is to find $\mathcal{G}^*$ that maximizes expected quality $Q(\cdot)$ without changing $f_{\text{ret}},f_{\text{sel}}$:
\begin{equation}\label{eq:objective}
\mathcal{G}^* = \arg\max_{\mathcal{G}} \mathbb{E}_{q, \mathcal{I}_t}\big[\, Q(q, \mathcal{R}(q, \mathcal{I}_t; \mathcal{G})) \big],
\end{equation}
subject to every $g\in\mathcal{G}$ encoding only how to decide from inventory evidence, never asserting a specific item's presence. $Q$ is a task-specific metric (e.g., satisfaction, conversion).

\subsection{Inventory-Grounded Online Search}
\label{sec:online}

\textbf{Inventory grounding.}
Inventory grounding probes the current inventory then fuses the results with pre-computed scene profiles into the \emph{inventory portrait} $P(q,\mathcal{I}_t)$.

\emph{Inventory probing} retrieves the top-$k_p$ matching items via fast embedding search:
\begin{equation}\label{eq:probe}
\mathcal{P} = \big\{ i \in \mathcal{I}_t \;\big|\; \operatorname{sim}(e_q, e_i) > \theta \big\}_{[k_p]},
\end{equation}
where $\operatorname{sim}$ is cosine similarity and $\operatorname{Cats}(\mathcal{P})$ denotes the unique categories among the probed items. Let $S_{\mathcal{P}} = \{\operatorname{sim}(e_q, e_i) \mid i \in \mathcal{P}\}$. Probe confidence summarizes the score distribution:
\begin{equation}\label{eq:probe-confidence}
\operatorname{conf}_p = \big(\max(S_{\mathcal{P}}),\; \overline{S_{\mathcal{P}}},\; \sigma(S_{\mathcal{P}})\big).
\end{equation}
For each $c \in \operatorname{Cats}(\mathcal{P})$, probe density counts how many probed items belong to $c$, and probe-level field coverage records field presence and fill rates within that category's probe subset:
\begin{equation}\label{eq:probe-signals}
\begin{aligned}
\mathcal{P}_c &= \{i \in \mathcal{P} : \operatorname{cat}(i) = c\},\\
\operatorname{density}_c &= |\mathcal{P}_c|,\\
(\operatorname{fields}_c^{\text{probe}},\; \operatorname{rel}_c^{\text{probe}}) &=
\operatorname{FieldStats}(\mathcal{P}_c).
\end{aligned}
\end{equation}
$\operatorname{conf}_p$ guides retrieval breadth, $\operatorname{density}_c$ steers search budget toward fertile categories, and $\operatorname{fields}_c^{\text{probe}}$, $\operatorname{rel}_c^{\text{probe}}$ reveal which fields are actually populated among items the query reaches.

\emph{Scene profile lookup} fetches pre-computed per-category metadata accumulated by a periodic offline pipeline that tracks recent inventory changes and known coverage gaps from validated historical failures:
\begin{equation}\label{eq:scene-profile}
\operatorname{SP}(c) = (\operatorname{cov\_gaps}_c,\; \operatorname{recent\_changes}_c).
\end{equation}
Scene profiles share the same scene-tag index as Policy Guidelines (Section~\ref{sec:offline}).

The inventory portrait templates together the probe confidence, per-category probe signals, and scene profiles:
\begin{multline}\label{eq:portrait}
P(q,\mathcal{I}_t) = \operatorname{Template}\bigl(
    \operatorname{conf}_p,\; \\
    \{(\mathcal{S}_c,\; \operatorname{SP}(c))
    \mid c \in \operatorname{Cats}(\mathcal{P}) \}\bigr),
\end{multline}
and $\mathcal{S}_c = (\operatorname{density}_c,\; \operatorname{fields}_c^{\text{probe}},\; \operatorname{rel}_c^{\text{probe}})$. Probe signals capture query-specific freshness; scene profiles contribute cross-query environment patterns. Only their fusion yields a complete inventory-grounded view.

\textbf{Guideline injection.} Each guideline $g$ is indexed by a scene tag (i.e., the category it applies to) and a stage tag $\tau\in\{\text{ret},\text{sel}\}$. Given the portrait's categories \(\operatorname{Cats}(P)\), we retrieve stage-matched candidates via an inverted index and independently select the top-\(k_g\) Guidelines for each stage by query-guideline embedding similarity:
\begin{equation}\label{eq:guideline-select}
\begin{aligned}
\mathcal{C}_{\tau}
&= \bigcup_{c\in\operatorname{Cats}(P)}
   \operatorname{Idx}_{\text{scene},\tau}^{-1}(c), \\
\mathcal{G}_{\tau}
&= \operatorname{TopK}_{k_g}\!\left(
   \mathcal{C}_{\tau};\operatorname{sim}(e_q,e_g)\right).
\end{aligned}
\end{equation}
where \(\mathcal{C}_{\tau}\) contains Guidelines matching both the portrait's scenes and stage \(\tau\), and \(\operatorname{TopK}_{k}\) returns the \(k\) highest-scoring elements. \(\mathcal{G}_{\text{ret}}\) and \(\mathcal{G}_{\text{sel}}\) are then injected into the corresponding prompts.

\subsection{Training-Free Policy-Level Optimization}
\label{sec:offline}
The offline optimizer updates $\mathcal{G}$ over $T$ rounds using a fixed inventory snapshot $\mathcal{I}$, without touching model weights or the retriever. Each round consists of trajectory grouping, inventory-guided exploration, and validated update.

\textbf{Rollout Classification and Trajectory Evidence.}
In each round, we sample queries from two sources: recent production logs and inventory-constructed probes that target category coverage gaps. We draw $N$ stochastic rollouts per query. A DeepSeek-V3 671B Judge (temperature 0.3) labels each rollout as \emph{success}, \emph{failure}, or \emph{uncertain}. Failure and uncertain outcomes are eligible for exploration, but uncertain outcomes are not used as negative evidence. A Query supplies contrastive evidence only when it has both a successful path---from an original rollout or a success recovered through exploration---and a failed rollout.

\textbf{Inventory-Guided Exploration Loop.}
For a Query \(q\) routed to exploration, we iteratively re-query the Retrieval Planner to search for missed routes using the current inventory evidence. In each turn, the Planner reads the inventory portrait $P(q,\mathcal{I})$ alongside accumulated observations from prior turns, identifies which portrait-described categories remain unexplored, and generates retrieval routes targeting those gaps. Starting from $\mathcal{O}_0=\emptyset$, at turn $k\ge1$,
\begin{equation}\label{eq:explore-step}
\mathcal{C}_k = f_{\text{ret}}\big(q,\; P(q,\mathcal{I}),\; \mathcal{G}_{\text{ret}},\; \operatorname{summary}(\mathcal{O}_{k-1})\big),
\end{equation}
where $\operatorname{summary}$ retains item names, categories, and key attributes. New candidates are merged: $\mathcal{O}_k = \mathcal{O}_{k-1}\cup\mathcal{C}_k$. The loop stops when accumulated items satisfy the request (a confirmed missed route that yields contrastive evidence), when portrait-covered categories are exhausted, or after \(K_{\max}\) turns.

\textbf{Guideline Update and Validation.}
Original or exploration-recovered successes are paired with failed rollouts from the same Query as contrastive evidence. DeepSeek-V3 671B (temperature 0.7) proposes and, when needed, consolidates scene-level candidate Patches from same-scene summaries of core requirements, relevant inventory evidence, and successful-versus-failed trajectory differences, together with deployed Guidelines and rejected-Patch summaries. The model selects JSON operations to add, revise, merge, or delete scene Guidelines; code attaches the fixed scene tag, validates the candidate JSON, and applies valid candidates to a temporary store for replay. Cases unresolved after finite inventory-guided exploration yield no positive evidence and do not proceed to Guideline proposal.

Let the proposed Patch be $\Delta\mathcal{G}$, and $\mathcal{G}' = \mathcal{G}\oplus\Delta\mathcal{G}$ (applying add, revise, merge, or delete operations). Code-valid candidate Patches first enter individual replay, where every Target Query is checked separately and all applicable Background and fixed-set regression gates must pass. Passing Patches are then added in ranked order to a provisional combination. After each addition, the resulting combination is checked against the same replay gates; a Patch that causes any gate to fail is removed, and only the final retained combination updates \(\mathcal{G}\). No Patch receives individual human approval. These finite replay gates provide empirical regression-risk control on the evaluated Queries; full gates, thresholds, and validation-set roles are provided in Appendix~\ref{sec:production-update-validation}.

\section{Experiments}
\label{sec:evaluation}

\begin{table*}[t]
\centering
\small
\setlength{\tabcolsep}{4.5pt}
\renewcommand{\arraystretch}{1.08}
\begin{tabular*}{0.94\textwidth}{@{\extracolsep{\fill}}llcccc@{}}
\toprule
\textbf{Setting} & \textbf{Method} & \textbf{Cand. R@30} & \textbf{F1@8} & \textbf{FNI (\%)} & \textbf{FM (\%)} \\
\midrule
\multicolumn{6}{@{}l}{\emph{Native configurations (single-run results)}} \\
\multirow{2}{*}{Native} & Production baseline & 0.691 & 0.643 & 30.6 & 60.8 \\
& \method{} (DeepSeek-V3) & \textbf{0.847} & \textbf{0.816} & \textbf{4.8} & \textbf{15.2} \\
\midrule
\multicolumn{6}{@{}l}{\emph{Token-matched prompt optimization: DeepSeek-V3 (3-run mean $\pm$ sample SD)}} \\
\multirow{2}{*}{18M tokens} & GEPA-style & 0.790 $\pm$ 0.004 & 0.787 $\pm$ 0.004 & 14.9 $\pm$ 0.9 & 46.0 $\pm$ 1.9 \\
& \method{} & \textbf{0.819 $\pm$ 0.007} & 0.792 $\pm$ 0.005 & \textbf{14.1 $\pm$ 1.1} & \textbf{34.7 $\pm$ 1.4} \\
\addlinespace[1pt]
\multirow{2}{*}{120M tokens} & GEPA-style & 0.823 $\pm$ 0.007 & 0.791 $\pm$ 0.006 & 12.7 $\pm$ 1.0 & 37.3 $\pm$ 1.8 \\
& \method{} & \textbf{0.850 $\pm$ 0.003} & \textbf{0.817 $\pm$ 0.004} & \textbf{4.9 $\pm$ 0.3} & \textbf{15.6 $\pm$ 0.9} \\
\midrule
\multicolumn{6}{@{}l}{\emph{Common Qwen3-8B policy backbone (3-run mean $\pm$ sample SD)}} \\
\multirow{4}{*}{Full pipeline} & SFT & 0.793 $\pm$ 0.004 & 0.763 $\pm$ 0.004 & 9.4 $\pm$ 0.7 & 28.2 $\pm$ 1.8 \\
& SFT+GRPO (shared reward) & 0.750 $\pm$ 0.010 & 0.699 $\pm$ 0.009 & 11.9 $\pm$ 1.0 & 39.4 $\pm$ 2.3 \\
& SFT+GRPO (stage-specific) & 0.796 $\pm$ 0.007 & 0.750 $\pm$ 0.006 & 8.5 $\pm$ 0.8 & 31.0 $\pm$ 1.6 \\
& \method{} & \textbf{0.815 $\pm$ 0.006} & \textbf{0.781 $\pm$ 0.008} & \textbf{6.2 $\pm$ 0.8} & \textbf{19.6 $\pm$ 1.7} \\
\bottomrule
\end{tabular*}
\caption{Main offline comparisons. The native-configuration block reports single-run results; the controlled blocks report three-run mean $\pm$ sample SD on the same 3,000-query test set. FNI and FM are percentages, with SDs in percentage points. Complete native-configuration metrics appear in Appendix Table~\ref{tab:native-results}.}
\label{tab:main-results}
\end{table*}

\subsection{Datasets}
\label{sec:datasets}

We use three internal datasets: 500 optimization queries drawn from online traffic and constructed exploratory cases, 500 disjoint held-out queries for validation replay, and 3,000 temporally later queries for final offline evaluation. The 500-query optimization set is used for \method{} and prompt/workflow optimization; weight-updating baselines additionally use a temporally earlier large-scale training corpus detailed in Appendix~\ref{sec:appendix}. All reported offline relevance and inventory-boundary metrics use expert annotations produced independently of \method{}'s exploration loop by product specialists and inventory-operations experts against the inventory snapshot used for evaluation. Optimization and validation replay use the same frozen inventory snapshot \(\mathcal{I}_t\), while the production retrieval interface remains fixed across all offline experiments. After validation, we freeze the selected Guideline store and evaluate it on the 3,000 later queries using snapshot \(\mathcal{I}_{t+1}\), taken four weeks after \(\mathcal{I}_t\), without further Patch proposal, validation, or Guideline update. Online serving uses current inventory evidence.

\subsection{Evaluation}

All systems use the same top-30 candidate and top-8 final-display budgets. For supported queries, annotators label a finite acceptable-target pool from candidates produced by evaluated systems and production retrieval logs; this pool is not an exhaustive list of all relevant inventory. Candidate Recall@30 measures target coverage before final selection. Precision@8, Recall@8, and F1@8 evaluate displayed results, with F1@8 as the primary offline metric.

We also report two boundary errors. False no-inventory rate is the fraction of supported queries for which a system incorrectly abstains or takes the no-inventory path. False match rate is the fraction of inventory-absence queries for which a system returns an inventory item as a match; abstentions, fallbacks, and explicit non-matching alternatives are counted as correct absence handling.

\subsection{Offline Experiments}

\subsubsection{Main Results}

Table~\ref{tab:main-results} separates three comparison settings. All rows are evaluated on the same 3,000-query final-test set at \(\mathcal{I}_{t+1}\), with the production retrieval interface and top-30/top-8 limits fixed. The native-configuration block reports single-run results, whereas the two controlled blocks report three-run mean $\pm$ sample SD. Baseline configurations and training details are provided in Appendix~\ref{sec:appendix}.

In the native single-run comparison, \method{} improves Candidate Recall@30 from 0.691 to 0.847 and F1@8 from 0.643 to 0.816 over the production baseline, while reducing false no-inventory from 30.6\% to 4.8\% and false match from 60.8\% to 15.2\%. This block describes end-to-end performance under the native configurations; the controlled comparisons below provide the basis for comparisons across optimization methods.

Under the token-matched DeepSeek-V3 setting, \method{} and the GEPA-style optimizer have comparable F1@8 at a budget of approximately 18M tokens ($0.792\pm0.005$ vs.\ $0.787\pm0.004$), with the mean difference falling within the reported run-to-run variation. At this budget, the clearer advantage of \method{} is its lower false-match rate (34.7\% vs.\ 46.0\%), alongside higher Candidate Recall@30 and a small reduction in false no-inventory. At a budget of approximately 120M tokens, \method{} has better mean values on all four metrics, increasing Candidate Recall@30 from 0.823 to 0.850 and F1@8 from 0.791 to 0.817, while reducing false no-inventory from 12.7\% to 4.9\% and false match from 37.3\% to 15.6\%.

With a common Qwen3-8B serving-policy backbone, \method{} obtains the highest mean Candidate Recall@30 and F1@8 and the lowest mean false no-inventory and false-match rates among the evaluated full-pipeline configurations. The stage-specific SFT+GRPO reward improves all four mean metrics over the shared reward, although its F1@8 remains below that of SFT (0.750 vs.\ 0.763). In this comparison, Qwen3-8B is used for all Retrieval Planner and Selection calls; \method{} uses DeepSeek-V3 only for offline judging and Patch generation.

\noindent\textbf{Optimization and serving footprint.}
\method{} runs three offline update rounds with $N=5$ rollouts per optimization query. Exploration considers at most $K_{\max}=10$ variants and occurs only during offline optimization; candidate Patches are applied only after validation replay. At serving time, the store is capped at 100 Guidelines, and the selector renders at most three Guidelines per stage, averaging 1.7 on covered requests. Added prompt length is 247.3 tokens on average and 568.2 at p95; latency overhead is 49.4 ms at p50 and 185.8 ms at p95. During the reported deployment period, replay-validated Guideline updates ran twice weekly in overnight batches, and whole-Patch rollback was supported. Optimization and serving details are given in Appendix~\ref{sec:appendix}; replay gates and observed production-update statistics are reported in Appendix~\ref{sec:production-update-validation}.

\noindent\textbf{Exploration recovery and audit.}
Across three offline optimization rounds, the exploration loop found successful retrieval routes for 112 of 124 inventory-supported cases in which the initial probe omitted the correct item (90.3\% cumulatively). Separately, among 93 query--snapshot cases unresolved after the three rounds, product and inventory-operations experts auditing the same frozen inventory snapshot identified 13 retrieval misses (14.0\%).

\subsubsection{Cross-Snapshot Analysis}

To examine performance under inventory evolution, we partition the 643 final-test Queries whose relevant inventory changed between $\mathcal{I}_t$ and $\mathcal{I}_{t+1}$ into five mutually exclusive change types. Table~\ref{tab:cross-snapshot-results} reports \method{} results from the single frozen-store run in the native-configuration block of Table~\ref{tab:main-results}; the selected Guideline store is not updated after the snapshot transition.

\begin{table}[t]
\centering
\small
\setlength{\tabcolsep}{3.5pt}
\begin{tabular}{@{}lrrrrr@{}}
\toprule
\textbf{Subset} & \textbf{Sup.} & \textbf{Abs.} & \textbf{F1@8} & \textbf{FNI} & \textbf{FM} \\
\midrule
\quad Additions & 275 & 52 & 0.797 & 8.7\% & 17.3\% \\
\quad Deletions & 35 & 51 & 0.761 & 11.4\% & 27.5\% \\
\quad Renames & 8 & 0 & 0.787 & 12.5\% & -- \\
\quad Metadata updates & 139 & 29 & 0.803 & 5.8\% & 24.1\% \\
\quad Schema changes & 43 & 11 & 0.780 & 9.3\% & 27.3\% \\
\midrule
All affected & 500 & 143 & 0.795 & 8.2\% & 23.1\% \\
Unaffected & 1,750 & 607 & 0.822 & 3.8\% & 13.3\% \\
\bottomrule
\end{tabular}
\caption{Cross-snapshot \method{} results. Affected categories are mutually exclusive. F1@8/FNI use supported Queries (Sup.); FM uses absence Queries (Abs.); a dash denotes no absence Query.}
\label{tab:cross-snapshot-results}
\end{table}

Across the five affected categories, F1@8 ranges from 0.761 to 0.803. On all affected Queries, \method{} achieves 0.795 F1@8 with an 8.2\% false-no-inventory rate on supported Queries and a 23.1\% false-match rate on absence Queries. The corresponding unaffected-query results are 0.822, 3.8\%, and 13.3\%, showing that snapshot-affected Queries form the more challenging portion of the final-test set.

\subsubsection{Ablation Study}

Table~\ref{tab:ablation-results} ablates stage-specific guideline placement, scene-conditioned selection, probe statistics, the inventory-guided exploration loop, and scene profiles. The stage-placement ablation uses one shared cross-stage guideline store for Retrieval and Summary.

\begin{table}[t]
\centering
\small
\setlength{\tabcolsep}{3pt}
\begin{tabular}{@{}p{0.42\linewidth}cccc@{}}
\toprule
\textbf{Variant} & \textbf{R@30} & \textbf{F1@8} & \textbf{FNI} & \textbf{FM} \\
\midrule
Shared cross-stage guidelines & 0.640 & 0.597 & 29.1\% & 57.6\% \\
w/o Scene selection & 0.812 & 0.783 & 8.0\% & 27.5\% \\
w/o Probe statistics & 0.804 & 0.776 & 17.6\% & 26.5\% \\
w/o Exploration loop & 0.811 & 0.789 & 11.6\% & 22.4\% \\
w/o Scene profiles & 0.838 & 0.798 & 8.8\% & 21.4\% \\
Full \method{} & \textbf{0.847} & \textbf{0.816} & \textbf{4.8\%} & \textbf{15.2\%} \\
\bottomrule
\end{tabular}
\caption{Ablation results. R@30: Candidate Recall@30; FNI/FM: false no-inventory/match rates.}
\label{tab:ablation-results}
\end{table}

The largest drop comes from shared cross-stage guidelines (F1@8 0.597), confirming that Retrieval and Selection need distinct guidance. Removing scene-conditioned selection (w/o Scene selection) increases FM to 27.5\%, since off-scene guidelines inject irrelevant constraints. Stripping probe statistics (w/o Probe statistics) degrades R@30 and FNI the most among inventory-grounded ablations: without confidence-graded routing signals, the method over-searches or under-searches. Dropping the exploration loop raises FNI and FM, as all-failure groups can no longer yield verified missed-route corrections. Removing scene profiles shows the smallest impact (R@30 0.838, F1@8 0.798), since probe statistics and the exploration loop partially compensate for missing coverage-gap and change-track signals.

\subsection{Online Experiments}

We run a fixed-duration 14-day online A/B test with 50/50 user-level randomization, covering 100,400 control users (249,100 sessions) and 99,800 treatment users (251,300 sessions). Both arms use the same Planner and Selection LLMs, base prompts, retrieval, ranking, and generation configuration. Control uses neither inventory portraits nor Guidelines, whereas treatment adds runtime inventory probing and portraits together with replay-validated Guidelines. The test therefore measures the end-to-end effect of the complete \method{} treatment rather than either component in isolation. We categorize traffic into exact, exploratory, composite, and inventory-absence slices (defined in Appendix~\ref{sec:appendix}); their traffic shares are 15.1\%, 37.8\%, 21.8\%, and 25.3\%, respectively. CTR lift is relative and estimated with user-clustered robust variance; the test has no significance-based early stopping. For the bad-case audit, we sample 500 Queries per arm in proportion to traffic shares and randomly within each arm-by-slice stratum. A product specialist and QA tester independently label each Query from its final results and contemporaneous inventory evidence while blinded to arm and internal method information; a senior product expert adjudicates disagreements.

\begin{table}[t]
\centering
\small
\begin{tabular}{@{}lcc@{}}
\toprule
\textbf{Slice} & \textbf{CTR Gain} & \textbf{Bad-case Red.} \\
\midrule
All queries & +3.17\% & 38.9\% \\
Exact & +0.97\% & 11.3\% \\
Exploratory & +4.27\% & 20.8\% \\
Composite & +7.19\% & 47.9\% \\
Inventory-absence & -- & 68.1\% \\
\bottomrule
\end{tabular}
\caption{Fourteen-day online A/B test. Bad-case reduction is relative to control over 1,000 audited Queries; for inventory-absence Queries, returning an item as a match counts as a bad case.}
\label{tab:online-results}
\end{table}

Table~\ref{tab:online-results} shows a 3.17\% relative CTR lift (95\% CI: 1.9--4.5\%; two-sided Wald $p<0.001$) and a 38.9\% reduction in audited bad cases (95\% CI: 25.2--50.1\%) over all traffic. Composite Queries have the largest observed CTR lift and bad-case reduction, while Exact Queries change less. The bad-case reduction intervals include zero reduction for Exact and Exploratory Queries but exclude it for Composite and inventory-absence Queries. For inventory-absence Queries, CTR is not reported because returning no match can be correct; the audited bad-case rate decreases from 45/128 in control to 14/125 in treatment.

\section{Conclusion}

We presented \method{}, a training-free approach to AI search over dynamic inventories. It grounds online decisions in inventory portraits and stage-indexed Policy Guidelines; offline rollouts and inventory-guided exploration support replay-validated updates. Offline replay and ablations improve candidate recall and F1@8; a 14-day online A/B test yields a 3.17\% relative CTR lift and 38.9\% fewer audited bad cases with bounded serving overhead. Deployed since May 2026, \method{} provides a practical way to adapt commercial AI search without retraining model weights.

\section{Limitations}

Our \method{} depends on the quality of runtime inventory evidence. Inventory portrait quality depends on retrieval probes, item metadata, and the production retrieval interface — each of which can be incomplete or noisy. When inventory fields are missing, scene coverage is sparse, or the retriever cannot expose relevant items through any available route, the optimizer may still struggle to distinguish a missed route from a lack of matching support in the observable inventory. The method reduces this ambiguity but cannot recover inventory or metadata that is not represented in the underlying system.

Policy Guidelines introduce maintenance overhead as inventory items, providers, and catalog schemas change. Our maintenance evidence covers a 31-day observation window at a 43--64-Guideline scale (Appendix~\ref{sec:operational-update-statistics}); longer-term growth, cross-system evaluation, full taxonomy migration, and proactive store-wide detection of stale or over-broad Guidelines remain outside the evaluated scope.

\section{Ethical Considerations}

The offline traces and online audits described in this paper are derived from production-style search interactions. All logged queries and audit records were anonymized and stripped of personally identifiable information before analysis. Data handling followed internal retention and access-control policies. Human auditors only saw information needed to judge search quality, and released examples should be anonymized or synthetic when raw queries or inventory items could reveal users, providers, or proprietary business information. The online A/B test was conducted through the production experimentation platform with standard monitoring and rollback controls.

\bibliography{references}

@inproceedings{lewis2020rag,
  title = {Retrieval-Augmented Generation for Knowledge-Intensive {NLP} Tasks},
  author = {Lewis, Patrick and Perez, Ethan and Piktus, Aleksandra and Petroni, Fabio and Karpukhin, Vladimir and Goyal, Naman and K{\"u}ttler, Heinrich and Lewis, Mike and Yih, Wen-tau and Rockt{\"a}schel, Tim and Riedel, Sebastian and Kiela, Douwe},
  booktitle = {Advances in Neural Information Processing Systems},
  volume = {33},
  pages = {9459--9474},
  publisher = {Curran Associates, Inc.},
  year = {2020},
  url = {https://proceedings.neurips.cc/paper_files/paper/2020/file/6b493230205f780e1bc26945df7481e5-Paper.pdf}
}

@inproceedings{karpukhin2020dpr,
  title = {Dense Passage Retrieval for Open-Domain Question Answering},
  author = {Karpukhin, Vladimir and Oguz, Barlas and Min, Sewon and Lewis, Patrick and Wu, Ledell and Edunov, Sergey and Chen, Danqi and Yih, Wen-tau},
  booktitle = {Proceedings of the 2020 Conference on Empirical Methods in Natural Language Processing},
  year = {2020},
  pages = {6769--6781},
  doi = {10.18653/v1/2020.emnlp-main.550},
  url = {https://aclanthology.org/2020.emnlp-main.550/}
}

@inproceedings{asai2024selfrag,
  title = {{SELF-RAG}: Learning to Retrieve, Generate, and Critique through Self-Reflection},
  author = {Asai, Akari and Wu, Zeqiu and Wang, Yizhong and Sil, Avirup and Hajishirzi, Hannaneh},
  booktitle = {International Conference on Learning Representations},
  year = {2024},
  url = {https://openreview.net/forum?id=hSyW5go0v8}
}

@inproceedings{yao2023react,
  title = {{ReAct}: Synergizing Reasoning and Acting in Language Models},
  author = {Yao, Shunyu and Zhao, Jeffrey and Yu, Dian and Du, Nan and Shafran, Izhak and Narasimhan, Karthik R. and Cao, Yuan},
  booktitle = {International Conference on Learning Representations},
  year = {2023},
  url = {https://openreview.net/forum?id=WE_vluYUL-X}
}

@article{shinn2023reflexion,
  title = {Reflexion: Language Agents with Verbal Reinforcement Learning},
  author = {Shinn, Noah and Cassano, Federico and Berman, Edward and Gopinath, Ashwin and Narasimhan, Karthik and Yao, Shunyu},
  journal = {arXiv preprint arXiv:2303.11366},
  year = {2023},
  doi = {10.48550/arXiv.2303.11366},
  url = {https://arxiv.org/abs/2303.11366}
}

@article{wang2026leaps,
  title = {{LEAPS}: An {LLM}-Empowered Adaptive Plugin in {Taobao} {AI} Search},
  author = {Wang, Lei and Wu, Jinhang and Wang, Zhibin and Li, Biye},
  journal = {arXiv preprint arXiv:2601.05513},
  year = {2026},
  url = {https://arxiv.org/abs/2601.05513}
}

@inproceedings{li2025searcho1,
  title = {Search-o1: Agentic Search-Enhanced Large Reasoning Models},
  author = {Li, Xiaoxi and Dong, Guanting and Jin, Jiajie and Zhang, Yuyao and Zhou, Yujia and Zhu, Yutao and Zhang, Peitian and Dou, Zhicheng},
  booktitle = {Proceedings of the 2025 Conference on Empirical Methods in Natural Language Processing},
  year = {2025},
  month = nov,
  address = {Suzhou, China},
  publisher = {Association for Computational Linguistics},
  pages = {5420--5438},
  doi = {10.18653/v1/2025.emnlp-main.276},
  url = {https://aclanthology.org/2025.emnlp-main.276/}
}

@inproceedings{nguyen2025minielm,
  title = {{MiniELM}: A Lightweight and Adaptive Query Rewriting Framework for {E}-Commerce Search Optimization},
  author = {Nguyen, Duy A. and Mohan, Rishi Kesav and Yang, Shimeng and Akash, Pritom Saha and Chang, Kevin Chen-Chuan},
  booktitle = {Findings of the Association for Computational Linguistics: ACL 2025},
  pages = {6952--6964},
  year = {2025},
  address = {Vienna, Austria},
  publisher = {Association for Computational Linguistics},
  doi = {10.18653/v1/2025.findings-acl.363},
  url = {https://aclanthology.org/2025.findings-acl.363/}
}

@inproceedings{lai2018productqa,
  title = {A Simple End-to-End Question Answering Model for Product Information},
  author = {Lai, Tuan and Bui, Trung and Li, Sheng and Lipka, Nedim},
  booktitle = {Proceedings of the First Workshop on Economics and Natural Language Processing},
  year = {2018},
  pages = {38--43},
  doi = {10.18653/v1/W18-3105},
  url = {https://aclanthology.org/W18-3105/}
}

@inproceedings{bagheri2022productrelevance,
  title = {Improving Relevance Quality in Product Search using High-Precision Query-Product Semantic Similarity},
  author = {Bagheri Garakani, Alireza and Yang, Fan and Hua, Wen-Yu and Chen, Yetian and Momma, Michinari and Deng, Jingyuan and Gao, Yan and Sun, Yi},
  booktitle = {Proceedings of the Fifth Workshop on e-Commerce and NLP},
  year = {2022},
  pages = {44--48},
  doi = {10.18653/v1/2022.ecnlp-1.6},
  url = {https://aclanthology.org/2022.ecnlp-1.6/}
}

@inproceedings{carbonell1998mmr,
  title = {The Use of {MMR}, Diversity-Based Reranking for Reordering Documents and Producing Summaries},
  author = {Carbonell, Jaime and Goldstein, Jade},
  booktitle = {Proceedings of the 21st Annual International ACM SIGIR Conference on Research and Development in Information Retrieval},
  year = {1998},
  pages = {335--336},
  doi = {10.1145/290941.291025},
  url = {https://doi.org/10.1145/290941.291025}
}

@inproceedings{santos2010xquad,
  title = {Exploiting Query Reformulations for Web Search Result Diversification},
  author = {Santos, Rodrygo L. T. and Macdonald, Craig and Ounis, Iadh},
  booktitle = {Proceedings of the 19th International Conference on World Wide Web},
  year = {2010},
  pages = {881--890},
  doi = {10.1145/1772690.1772780},
  url = {https://doi.org/10.1145/1772690.1772780}
}

@inproceedings{yang2024opro,
  title = {Large Language Models as Optimizers},
  author = {Yang, Chengrun and Wang, Xuezhi and Lu, Yifeng and Liu, Hanxiao and Le, Quoc V. and Zhou, Denny and Chen, Xinyun},
  booktitle = {International Conference on Learning Representations},
  year = {2024},
  url = {https://openreview.net/forum?id=Bb4VGOWELI}
}

@inproceedings{khattab2024dspy,
  title = {{DSPy}: Compiling Declarative Language Model Calls into State-of-the-Art Pipelines},
  author = {Khattab, Omar and Singhvi, Arnav and Maheshwari, Paridhi and Zhang, Zhiyuan and Santhanam, Keshav and Vardhamanan A, Sri and Haq, Saiful and Sharma, Ashutosh and Joshi, Thomas T. and Moazam, Hanna and Miller, Heather and Zaharia, Matei and Potts, Christopher},
  booktitle = {International Conference on Learning Representations},
  year = {2024},
  url = {https://openreview.net/forum?id=sY5N0zY5Od}
}

@inproceedings{agrawal2026gepa,
  title = {{GEPA}: Reflective Prompt Evolution Can Outperform Reinforcement Learning},
  author = {Agrawal, Lakshya A. and Tan, Shangyin and Soylu, Dilara and Ziems, Noah and Khare, Rishi and Opsahl-Ong, Krista and Singhvi, Arnav and Shandilya, Herumb and Ryan, Michael J. and Jiang, Meng and Potts, Christopher and Sen, Koushik and Dimakis, Alexandros G. and Stoica, Ion and Klein, Dan and Zaharia, Matei and Khattab, Omar},
  booktitle = {International Conference on Learning Representations},
  year = {2026},
  note = {Oral presentation},
  url = {https://openreview.net/forum?id=RQm2KQTM5r}
}

@article{shao2024deepseekmath,
  title = {{DeepSeekMath}: Pushing the Limits of Mathematical Reasoning in Open Language Models},
  author = {Shao, Zhihong and Wang, Peiyi and Zhu, Qihao and Xu, Runxin and Song, Junxiao and Bi, Xiao and Zhang, Haowei and Zhang, Mingchuan and Li, Y. K. and Wu, Y. and Guo, Daya},
  journal = {arXiv preprint arXiv:2402.03300},
  year = {2024},
  url = {https://arxiv.org/abs/2402.03300}
}

@inproceedings{elsas2010temporal,
  title = {Leveraging Temporal Dynamics of Document Content in Relevance Ranking},
  author = {Elsas, Jonathan L. and Dumais, Susan T.},
  booktitle = {Proceedings of the Third ACM International Conference on Web Search and Data Mining},
  series = {WSDM '10},
  year = {2010},
  pages = {1--10},
  doi = {10.1145/1718487.1718489},
  url = {https://doi.org/10.1145/1718487.1718489}
}

\appendix
\clearpage
\appendix
\twocolumn[
\begin{@twocolumnfalse}
\section{Additional Results and Experimental Details}
\label{sec:appendix}

\subsection{Native-Configuration Results}

\begin{center}
\centering
\small
\resizebox{\textwidth}{!}{%
\begin{tabular}{lcccccc}
\toprule
\textbf{Method} & \textbf{Cand. R@30} & \textbf{P@8} & \textbf{R@8} & \textbf{F1@8} & \textbf{False No-Inv.} & \textbf{False Match} \\
\midrule
\multicolumn{7}{l}{\emph{Production system}} \\
Production Baseline & 0.691 & 0.612 & 0.678 & 0.643 & 30.6\% & 60.8\% \\
\midrule
\multicolumn{7}{l}{\emph{Training-free / search-time alternatives}} \\
ReflectiveRAG-style Self-Reflective Retrieval & 0.766 & 0.753 & 0.742 & 0.747 & 14.8\% & 41.3\% \\
Reflexion-style Episodic Memory & 0.779 & 0.759 & 0.749 & 0.754 & 18.4\% & 48.0\% \\
ReAct-style Search Agent & 0.781 & 0.756 & 0.752 & 0.754 & 10.3\% & 43.5\% \\
\midrule
\multicolumn{7}{l}{\emph{Prompt/workflow optimization}} \\
GEPA-style Prompt Optimizer & 0.789 & 0.796 & 0.781 & 0.788 & 14.2\% & 46.0\% \\
\midrule
\multicolumn{7}{l}{\emph{Weight-updating alternatives}} \\
SFT Retrieval Planner (Qwen3-8B) & 0.793 & 0.734 & 0.754 & 0.744 & 11.0\% & 32.4\% \\
SFT+GRPO Retrieval Planner (Qwen3-8B) & 0.760 & 0.724 & 0.722 & 0.723 & 12.7\% & 36.0\% \\
SFT Full Pipeline (Qwen3-8B) & 0.795 & 0.752 & 0.775 & 0.763 & 9.7\% & 26.4\% \\
SFT+GRPO Full Pipeline (Qwen3-8B) & 0.744 & 0.718 & 0.689 & 0.703 & 11.5\% & 38.4\% \\
\midrule
\multicolumn{7}{l}{\emph{Ours}} \\
\method{} & \textbf{0.847} & \textbf{0.812} & \textbf{0.820} & \textbf{0.816} & \textbf{4.8\%} & \textbf{15.2\%} \\
\bottomrule
\end{tabular}
}
\captionsetup{hypcap=false}
\captionof{table}{Complete native-configuration results from the submitted single run. These systems use their originally evaluated backbones and optimization budgets, so the table provides comparison breadth rather than a controlled comparison of optimization methods.}
\label{tab:native-results}
\end{center}
\end{@twocolumnfalse}
]

Table~\ref{tab:native-results} reports the complete single-run native-configuration comparison. Its Production Baseline and \method{} rows correspond to the native-configuration block in Table~\ref{tab:main-results}; the controlled blocks in Table~\ref{tab:main-results} report separate three-run comparisons.

\subsection{Baseline Configurations}

The ReflectiveRAG-style, Reflexion-style, ReAct-style, and GEPA-style baselines use the same frozen DeepSeek-V3 backbone, production retrieval interface, and inventory portrait as the DeepSeek-V3 configuration of \method{}. They differ in how they orchestrate retrieval actions, use failure feedback, or optimize the two-stage prompts.

\textbf{ReflectiveRAG-style Self-Reflective Retrieval.}
A lightweight reflection controller, also DeepSeek-V3, iteratively evaluates evidence sufficiency after each retrieval round and adaptively reformulates queries when evidence is incomplete, up to 3 reflection rounds.

\textbf{Reflexion-style Episodic Memory.}
The agent maintains an episodic memory of natural-language reflections from prior failures. At each trial, it retrieves and reasons with the current query, memory, and inventory portrait, up to 5 trials per query.

\textbf{ReAct-style Search Agent.}
The agent interleaves reasoning steps with retrieval tool calls, observing intermediate results before taking the next action, up to 10 interaction steps per query. Tool calls query the same production retrieval interface.

\textbf{GEPA-style Prompt Optimizer.}
Following the GEPA framework, a DeepSeek-V3 optimizer reflects on trajectory traces to propose prompt revisions for both the Retrieval Planner and Selection stages. In the native configuration reported in Table~\ref{tab:native-results}, optimization uses a total rollout budget of 1,000, a minibatch of 16 queries per proposal, and Pareto-based candidate selection. We adapt GEPA's reflective prompt mutation to the two-stage pipeline by alternating module updates in round-robin order. The token-matched runs in Table~\ref{tab:main-results} use the same implementation and stop at approximately 18M or 120M total input-plus-output tokens. When the next complete optimization operation would exceed the budget, the run stops and retains the best fully evaluated prompt candidate.

\subsection{Training Details for Weight-Updating Baselines}

Both Retrieval-Planner-only and full-pipeline variants of SFT and SFT+GRPO use Qwen3-8B as the student backbone. Training data consists of 100k query--target pairs drawn from production search logs and augmented with synthetic pairs generated by a DeepSeek-V3 teacher, following the inverse data augmentation strategy of LEAPS \citep{wang2026leaps}. SFT uses standard next-token cross-entropy for 2 epochs with the AdamW optimizer, a learning rate of $2{\times}10^{-5}$, batch size 128, and a cosine learning-rate schedule with linear warmup. GRPO further trains the SFT checkpoint on a 5k-query high-variance subset with group size $G{=}8$ and learning rate $1{\times}10^{-6}$ on 8 NVIDIA H800 GPUs. Let $r_{\mathrm{retrieval}}$ denote Candidate Recall@30, and let $r_{\mathrm{outcome}}$ be 1 when the final results satisfy the Query or correctly handle inventory absence, 0 when uncertain, and $-1$ for a core-constraint violation, false no-inventory, or false match. The SFT+GRPO rows in Table~\ref{tab:native-results} use the submitted reward design: the Retrieval-Planner-only variant uses $r_{\mathrm{retrieval}}$ for supported Queries and $r_{\mathrm{outcome}}$ for inventory-absence Queries, while the Full-Pipeline variant uses $R=0.5r_{\mathrm{retrieval}}+0.5r_{\mathrm{outcome}}$ for supported Queries and $R=r_{\mathrm{outcome}}$ otherwise. This rollout-level scalar updates both stages in the Full-Pipeline variant. In the stage-specific configuration in Table~\ref{tab:main-results}, the Planner receives $r_{\mathrm{retrieval}}$ and Selection receives $r_{\mathrm{outcome}}$. The shared- and stage-specific controlled configurations use the same SFT initialization, training data, settings, and random seeds.

\subsection{Optimization and Serving Configuration}

For each optimization Query, \method{} draws $N{=}5$ stochastic rollouts. The exploration loop (\S\ref{sec:offline}) runs for at most $K_{\max}{=}10$ turns per Query and stops earlier when accumulated items satisfy the request or all portrait-covered categories have been explored. Model and temperature settings for rollout classification, Query-level summarization, Patch proposal and consolidation, and pairwise replay are reported with the prompt templates in Appendix~\ref{sec:prompt-templates}. At serving time, the selector first filters Guidelines by scene and stage using the inverted index and then independently selects at most three for each stage by query--Guideline embedding similarity, as defined in~\S\ref{sec:online}. The Guideline store is capped at 100 entries.

\subsection{Online Traffic Slices}

Exact queries specify a concrete item or service with directly matchable attributes. Exploratory queries describe an underspecified need or scenario that requires inventory-grounded expansion. Composite queries combine multiple constraints, scenarios, or service requirements and usually require multi-route retrieval and final selection. Inventory-absence queries have no matching support under the observed inventory evidence.

\section{Production Update Validation and Maintenance}
\label{sec:production-update-validation}

\subsection{Patch Replay Validation}

\noindent\textbf{Pre-replay checks.} Code first checks every proposed Patch before replay. Its JSON must satisfy the schema; every referenced Guideline ID, edit operation, and scene/stage tag must be valid; and the Patch must contain no code-detectable conflict, such as revising and deleting the same Guideline. Candidates that pass enter individual replay, one Patch at a time.

\noindent\textbf{Individual replay.} This stage compares the deployed Guideline store fixed at the start of the update batch (\emph{Current}) with the store obtained by applying one candidate Patch (\emph{Candidate}). The two stores are run under up to three matched random seeds, meaning that both use the same seed in each paired trial. We evaluate two Query groups. Target Queries are the failure cases that motivated the Patch and that it aims to repair. Background Queries are production-log or inventory-constructed probe Queries for which Current had previously produced successful outputs; they test whether the Patch disrupts cases that already worked. For both groups, a pairwise Judge compares Candidate and Current and returns a win when Candidate is better, a loss when Current is better, or uncertain when neither can be reliably preferred. Separately, a labeled operational regression set fixed within each replay batch (545 Queries in the reported workflow) measures Candidate's Precision, Recall, and F1 drops relative to Current. For the offline evaluation reported in this paper, the 500 held-out Queries in Section~\ref{sec:datasets} form the validation set used to select the Guideline store on \(\mathcal{I}_t\); routine production Patch acceptance instead uses the operational regression set described above.

Individual replay accepts a Patch only if every Target Query passes the win/loss and uncertainty gates and the Background and fixed-set checks satisfy all applicable aggregate and per-seed thresholds in Table~\ref{tab:patch-gates}. Replay may stop early only when a candidate can no longer pass even if all remaining comparisons are favorable; every accepted candidate therefore completes all three seeds. Because target gates apply separately, a strong repair on one Target Query cannot offset failure on another. Aggregate gates assess performance across all three seeds, while the looser per-seed caps prevent a severe regression in one seed from being hidden by aggregation.

\begin{table}[!ht]
\centering
\small
\setlength{\tabcolsep}{3pt}
\begin{tabular*}{\linewidth}{@{\extracolsep{\fill}}lcc@{}}
\toprule
\textbf{Gate} & \textbf{Aggregate gate} & \textbf{Per-seed cap} \\
\midrule
Target win/loss & Wins $>$ losses & -- \\
Target uncertain & $\leq1$ of 3 & -- \\
Background loss rate & $\leq2.0\%$ & $\leq4.0\%$ \\
Background uncertain rate & $\leq5.0\%$ & $\leq10.0\%$ \\
Precision drop & $\leq0.5$ pp & $\leq1.0$ pp \\
Recall drop & $\leq0.5$ pp & $\leq1.0$ pp \\
F1 drop & $\leq0.3$ pp & $\leq0.6$ pp \\
\bottomrule
\end{tabular*}
\caption{Patch replay acceptance thresholds. Target gates use three paired trials per Target Query. Background rates pool all Query--seed judgments, while metric drops are computed per seed on the labeled 545-Query operational set and then averaged equally across seeds. Values equal to a listed $\leq$ threshold are accepted.}
\label{tab:patch-gates}
\end{table}

\noindent\textbf{Combined replay.} Patches that pass individual replay are ranked using the following priorities, in order: (i) higher mean Target net win rate, computed as $(\text{wins}-\text{losses})/3$ for each Target Query and then averaged across Target Queries; (ii) higher mean F1 gain relative to Current; (iii) lower Background loss rate; (iv) lower Background uncertain rate; and (v) earlier proposal order. The ordered Patches are added one at a time to a provisional combination. Each resulting combination is checked against the same batch-start Current using the same gates, and every retained combination completes three full 545-Query replays. If adding a Patch causes the resulting combination to fail any gate, that Patch is removed and the previously retained combination is kept. After all Patches have been considered, the final retained combination becomes the deployed store. These finite checks provide empirical regression-risk control on the evaluated Queries.

\subsection{Operational Update Statistics}
\label{sec:operational-update-statistics}

During May 15--June 14 (31 days), nine automatic Guideline-update batches ran overnight, twice weekly. Across these batches, 83 candidate Patches entered replay validation, and 45 passed both individual and final combined replay and were applied (54.2\%). Table~\ref{tab:patch-operation-statistics} reports the corresponding operation counts. Operation totals exceed Patch totals because a Patch may contain multiple operations.

Over the same period, the Guideline store grew from 43 to 62 entries and peaked at 64. Update runs took a median 2.2 hours (range: 1.7--3.2), averaging 10.8K LLM calls and 30.8M tokens per batch; incremental maintenance averaged 0.7 person-hours per week. A post-hoc audit of the 19 accepted revise operations and one accepted delete operation identified four lifecycle cases---one stale and three over-broad Guidelines---addressed by three revises and one delete. A separate audit found no Guideline assigned to the wrong Scene tag; Query-level routing errors were not measured.

\begin{table}[!ht]
\centering
\small
\setlength{\tabcolsep}{5pt}
\begin{tabular}{@{}lrr@{}}
\toprule
\textbf{Operation} & \textbf{Proposed} & \textbf{Accepted/applied} \\
\midrule
Add & 49 & 25 \\
Revise & 27 & 19 \\
Merge & 12 & 5 \\
Delete & 3 & 1 \\
\bottomrule
\end{tabular}
\caption{Patch operation statistics over nine production update batches. Proposed operations belong to Patches that entered replay validation; accepted/applied operations belong to Patches retained after individual and combined replay.}
\label{tab:patch-operation-statistics}
\end{table}

\section{Prompt Templates}
\label{sec:prompt-templates}

\noindent\textbf{Provenance and configuration.}
The production workflow used Chinese prompts. We provide anonymized English versions that preserve the task definitions, decision rules, inputs, limits, and output schemas while omitting product-specific identifiers and non-functional platform wrappers. All five calls use DeepSeek-V3 671B. Rollout classification, Query-level summarization, and pairwise replay use temperature 0.3; Patch proposal and scene-level consolidation use temperature 0.7. Calling code supplies at most eight same-scene Query summaries to one proposer call. A proposer returns at most four Patches with at most three operations per Patch; consolidation uses the same limits.

\subsection{Patch Proposer}

\begin{lstlisting}[style=prompt,caption={Anonymized English version of the Patch-proposer prompt.},label={prompt:patch-proposer}]
TASK
You are a search-rule optimization expert. Based on summaries of
multiple search Queries in the current scene, the currently deployed
Guidelines, and previously rejected changes, generate candidate
Guideline Patches.

TERMINOLOGY
- retrieval stage: Given a Query and runtime inventory evidence,
  generate and execute search routes, including the search object,
  fields, keywords, and constraints, to form a candidate item set.
- selection stage: Filter the candidate item set against the
  Query's core intent and constraints to produce the final results.
- Guideline: A rule that directs retrieval- or selection-stage
  behavior.
- Patch: A set of candidate Guideline changes for one reusable
  failure pattern. A Patch may contain one or more add, revise,
  merge, or delete operations.

LIMITS
- The patches list may contain at most {max_patches} Patches.
- Each Patch may contain at most {max_operations_per_patch}
  operations.

INPUTS

1. Current scene tag
{scene_tag}

2. Query-level summaries
The summaries are provided as a list. Each record contains:
- query_id: the unique identifier of the Query;
- query_text: the original Query text;
- core_requirements: the explicitly stated core intent, object,
  attributes, and constraints;
- inventory_portrait_summary: a summary of inventory categories,
  fields, coverage, and change signals relevant to the Query;
- successful_patterns: retrieval or selection approaches used by
  successful trajectories or successful paths recovered through
  exploration;
- failed_patterns: omissions, violations, or incorrect handling in
  clearly failed trajectories; and
- key_differences: outcome-relevant differences between successful
  and failed paths, including evidence that the failure occurred in
  retrieval, selection, or both.

{query_summaries}

3. Currently deployed Guidelines for the scene
The Guidelines are provided as a list. Each record contains:
- guideline_id: the Guideline's fixed unique identifier;
- stage: retrieval or selection; and
- guideline_text: the complete text currently in effect.

{deployed_guidelines}

4. Summaries of previously rejected Patches for the scene
The rejected Patches are provided as a list of structured text
summaries. Each contains a heading with the Patch ID, rejection
stage, and scene, followed by Target, Attempt, Result, and Subsequent
Constraint fields. When available, Result includes the replay
failure reason and key metrics. Use this evidence to avoid repeating
changes already shown to be ineffective, over-broad, conflicting,
or regressive.

{rejected_patch_summaries}

PATCH GENERATION PROCEDURE
1. Inspect the evidence in every Query-level summary: the core
   requirements, inventory evidence, successful patterns, failed
   patterns, and key differences.
2. Identify failure patterns supported by one or more Queries and
   reusable for similar Queries. A failure pattern should describe a
   reusable retrieval or selection problem, not a fact that applies
   only to one item or one inventory snapshot.
3. Determine whether each pattern concerns retrieval, selection, or
   both. Check whether the deployed Guidelines already cover it and
   whether the issue is missing coverage, an incomplete condition,
   over-broad scope, ambiguous wording, an incorrect rule, overlap,
   or persistent harm.
4. Inspect the targets, attempted changes, replay failures, and key
   metrics in the rejected-Patch summaries. If current evidence
   resolves the earlier failure, a narrowed or corrected candidate
   may be proposed; explain the difference in reason.
5. Select add, revise, merge, or delete using the rules below. Keep
   each Patch local, clearly bounded, and as small as possible. When
   an existing Guideline can be modified, do not add a semantic
   duplicate.
6. Jointly inspect all Patches generated in this call. Resolve
   inconsistent edits to the same Guideline and exclude candidates
   that duplicate, contain, or semantically conflict with another.
7. Generate no Patch when evidence is insufficient, no reusable
   pattern exists, the deployed Guidelines already cover the change,
   or the only possible rule would encode a current inventory fact.

OPERATION FIELDS
Every operation contains:
- op: add, revise, merge, or delete;
- stage: retrieval or selection;
- target_guideline_ids: the deployed Guideline IDs affected by the
  operation, with cardinality determined by op;
- guideline_text: the complete Guideline after the operation, or
  null for delete; and
- reason: the supporting evidence and rationale for the operation.

OPERATION SELECTION RULES

add
Use add when no related deployed Guideline covers the failure pattern
and a new, reusable retrieval or candidate-selection method is needed.
- target_guideline_ids must be an empty list.
- guideline_text must contain the complete new Guideline.

revise
Use revise when one related Guideline has incomplete applicability
conditions, over-broad scope, ambiguous wording, or prescribes an
incorrect method.
- target_guideline_ids must contain exactly one deployed ID.
- guideline_text must contain the complete revised Guideline, not
  only the changed fragment.

merge
Use merge when at least two deployed Guidelines duplicate or
substantially overlap, or when a conflict can be resolved by one
formulation without losing valid conditions or methods.
- target_guideline_ids must contain at least two deployed IDs.
- guideline_text must contain the complete merged Guideline.
- Do not merge Guidelines with different applicability conditions or
  methods merely to reduce the number of Guidelines.

delete
Use delete when one Guideline persistently causes errors across
multiple related Queries or historical replay and has no useful part
to retain.
- target_guideline_ids must contain exactly one deployed ID.
- guideline_text must be null.
- If useful content remains, use revise or merge instead of delete.

GUIDELINE REQUIREMENTS
1. A Guideline should provide retrieval or candidate-selection methods
   reusable across similar Queries rather than restating the answer to
   one Query.
2. State an explicit applicability condition and an executable
   action, including how runtime inventory evidence should be used
   when relevant.
3. Do not encode item IDs, item names, current inventory counts, or
   claims that an item currently exists or is absent.
4. Do not turn transient category coverage, field fill, or
   recent-change signals from the inventory portrait into permanent
   inventory knowledge.
5. A retrieval Guideline should state when and how the Retrieval
   Planner should apply a retrieval method, such as how to choose
   search objects and fields, organize keywords and constraints, and
   construct retrieval routes.
6. A selection Guideline should state when and how to apply a
   candidate-selection method, such as how to check core requirements,
   reject weak matches, compare candidates, and organize the final
   results.
7. Use only evidence supplied in this prompt. Do not invent user
   requirements, inventory fields, business rules, or failure causes.
8. A Query without a verified successful path, including one for
   which finite exploration found no support, cannot by itself
   produce a positive Guideline.
9. Prefer local, clear, reviewable changes; do not rewrite Guidelines
   unrelated to the target failure pattern.

PATCH ORGANIZATION REQUIREMENTS
1. Each Patch addresses one target_failure_pattern.
2. Put all operations needed for the same failure pattern in one
   Patch; use separate Patches for different patterns.
3. Mark every operation as retrieval or selection.
4. Every Patch must list the input target_query_ids that directly
   support its failure pattern. Do not cite an ID absent from input.
5. revise, merge, and delete may reference only deployed Guideline
   IDs.
6. reason must identify the supporting evidence and explain the
   operation choice. When relevant, explain how the candidate differs
   from a related rejected Patch.

OUTPUT REQUIREMENTS
Return exactly one JSON object. Do not return Markdown, analysis, or
fields outside this schema:
{
  "patches": [
    {
      "target_failure_pattern": "<reusable failure pattern>",
      "target_query_ids": ["Q1", "Q2"],
      "operations": [
        {
          "op": "add",
          "stage": "retrieval",
          "target_guideline_ids": [],
          "guideline_text": "<complete new Guideline>",
          "reason": "<supporting evidence and operation rationale>"
        }
      ]
    }
  ],
  "no_patch_reason": null
}

NO-PATCH CASE
- If patches is nonempty, no_patch_reason must be null.
- If evidence does not support a Patch, return an empty patches list
  and briefly explain the reason in no_patch_reason.

OUTPUT DISCIPLINE
Use only the supplied evidence. Follow all operation rules, limits,
and output requirements. Do not return an unsupported Patch.
\end{lstlisting}

\noindent\textbf{Implementation note.}
The calling pipeline fixes the current scene tag. After deterministic pre-replay validation, code assigns each retained candidate a stable batch-scoped Patch ID; the LLM does not generate this operational identifier.

\subsection{Scene-Level Patch Consolidation}

This call occurs only when a scene requires multiple proposer minibatches; otherwise the single proposer output proceeds directly to deterministic validation.

\begin{lstlisting}[style=prompt,caption={Anonymized English version of the scene-level consolidation prompt.},label={prompt:patch-consolidation}]
TASK
You are a search-rule optimization expert. Organize candidate Patches
produced by multiple minibatches for the same scene. Using the
currently deployed Guidelines, combine duplicate or overlapping
candidates, handle conflicts, and return the final candidate Patches.

TERMINOLOGY
- retrieval stage: Given a Query and runtime inventory evidence,
  generate and execute search routes, including the search object,
  fields, keywords, and constraints, to form a candidate item set.
- selection stage: Filter the candidate item set against the
  Query's core intent and constraints to produce the final results.
- Guideline: A rule that directs retrieval- or selection-stage
  behavior.
- Patch: A set of candidate Guideline changes for one reusable
  failure pattern. A Patch may contain one or more add, revise,
  merge, or delete operations.

LIMITS
- The patches list may contain at most
  {max_consolidated_patches} Patches.
- Each Patch may contain at most {max_operations_per_patch}
  operations.

INPUTS

1. Currently deployed Guidelines for the scene
The Guidelines are provided as a list. Each record contains:
- guideline_id: the Guideline's fixed unique identifier;
- stage: retrieval or selection; and
- guideline_text: the complete text currently in effect.

{deployed_guidelines}

2. Candidate Patches produced by the minibatches
The candidates are provided as a list. Each contains:
- target_failure_pattern: the reusable failure pattern addressed;
- target_query_ids: Query IDs directly supporting that pattern; and
- operations: one or more candidate Guideline operations, each with
  op, stage, target_guideline_ids, guideline_text, and reason.

{candidate_patches}

PATCH CONSOLIDATION PROCEDURE
1. Inspect each candidate's failure pattern, supporting Queries,
   Guideline operations, and reasons. Compare target IDs, stages,
   applicability conditions, and proposed text with the deployed
   Guidelines.
2. If candidates address the same failure pattern with equivalent
   changes, combine them into one Patch. Deduplicate target_query_ids
   while retaining complete, clearly bounded Guideline text and
   operation rationale.
3. If candidates partially overlap, consolidate them only when doing
   so neither broadens applicability nor removes valid conditions.
   Otherwise keep them separate with non-overlapping operation
   boundaries; if that is impossible, treat them as conflicting.
4. Check for inconsistent edits to the same Guideline, simultaneous
   revision and deletion, contradictory Guideline text, or different
   operations that propose semantically equivalent Guideline changes.
   Resolve conflicts using only the input candidates and deployed
   Guidelines. Exclude a conflicting change that cannot be resolved
   from the inputs.
5. Ensure that each final Patch addresses one reusable pattern and
   that the final Patches do not duplicate, contain, conflict with,
   or inconsistently modify one another.

OPERATION FIELDS
Every operation contains:
- op: add, revise, merge, or delete;
- stage: retrieval or selection;
- target_guideline_ids: affected deployed Guideline IDs;
- guideline_text: the complete Guideline after the operation, or
  null for delete; and
- reason: supporting input candidates, the operation rationale, and
  how duplicate, overlapping, or conflicting candidates were handled.

OPERATION SELECTION RULES
Consolidating multiple candidate Patches is not itself the merge
operation. Select op according to the actual change applied to the
deployed Guidelines.

add
Use add when no deployed Guideline covers the failure pattern and a
new, reusable retrieval or candidate-selection method is needed.
- target_guideline_ids must be empty.
- guideline_text must contain the complete new Guideline.

revise
Use revise when one related Guideline has incomplete applicability
conditions, over-broad scope, ambiguous wording, or prescribes an
incorrect method.
- target_guideline_ids must contain exactly one deployed ID.
- guideline_text must contain the complete revised Guideline.

merge
Use merge when at least two deployed Guidelines duplicate or
substantially overlap, or when a conflict can be resolved by one
formulation without losing valid conditions or methods.
- target_guideline_ids must contain at least two deployed IDs.
- guideline_text must contain the complete merged Guideline.
- Do not merge Guidelines with different applicability conditions or
  methods merely to reduce the number of Guidelines.

delete
Use delete when the input reason establishes that one Guideline
persistently harms multiple related Queries and has no useful part to
retain.
- target_guideline_ids must contain exactly one deployed ID.
- guideline_text must be null.
- If useful content remains, use revise or merge instead of delete.

GUIDELINE REQUIREMENTS
1. A Guideline should provide retrieval or candidate-selection methods
   reusable across similar Queries rather than restating the answer to
   one Query.
2. Retain explicit applicability conditions and executable actions,
   including use of runtime inventory evidence when relevant.
3. Do not encode item IDs, item names, current inventory
   counts, current existence or absence claims, or transient category
   coverage, field fill, or recent-change signals.
4. A retrieval Guideline should state when and how the Retrieval
   Planner should apply a retrieval method, such as how to choose
   search objects and fields, organize keywords and constraints, and
   construct retrieval routes.
5. A selection Guideline should state when and how to apply a
   candidate-selection method, such as how to check core requirements,
   reject weak matches, compare candidates, and organize the final
   results.
6. Use only the input candidates and deployed Guidelines. Do not
   invent requirements, inventory fields, business rules, or failure
   causes.
7. Prefer local, clear, reviewable changes; do not rewrite Guidelines
   unrelated to the input candidates.

PATCH ORGANIZATION REQUIREMENTS
1. Each Patch addresses one target_failure_pattern.
2. Put all operations needed for one failure pattern in one Patch;
   use separate Patches for different patterns.
3. Mark every operation as retrieval or selection.
4. Deduplicate target_query_ids and cite only IDs found in the input.
5. revise, merge, and delete may reference only deployed IDs.
6. reason must identify the supporting candidates and explain how
   duplicate, overlapping, or conflicting candidates were handled.

OUTPUT REQUIREMENTS
Return exactly one JSON object. Do not return Markdown, analysis, or
fields outside this schema:
{
  "patches": [
    {
      "target_failure_pattern": "<reusable failure pattern>",
      "target_query_ids": ["Q1", "Q2"],
      "operations": [
        {
          "op": "add",
          "stage": "retrieval",
          "target_guideline_ids": [],
          "guideline_text": "<complete new Guideline>",
          "reason": "<supporting candidates and consolidation rationale>"
        }
      ]
    }
  ],
  "no_patch_reason": null
}

NO-PATCH CASE
- If patches is nonempty, no_patch_reason must be null.
- If deployed Guidelines already cover all candidates, or no remaining
  change can be resolved from the input, return an empty patches list
  and explain the reason in no_patch_reason.

OUTPUT DISCIPLINE
Use only the input candidates and deployed Guidelines. Follow all
operation rules, limits, and output requirements. Do not introduce an
unsupported change.
\end{lstlisting}

\noindent\textbf{Implementation note.}
As with proposer output, calling code assigns stable batch-scoped Patch IDs only after deterministic validation.

\subsection{Rollout Classification Judge}

\begin{lstlisting}[style=prompt,caption={Anonymized English version of the rollout-classification Judge prompt.},label={prompt:rollout-judge}]
TASK
You are a search-result quality reviewer. Given one search Query, its
inventory portrait, item evidence, and the final results from
multiple rollouts, independently classify every rollout as success,
failure, or uncertain.

INPUTS

1. Query
query_text: {query_text}

2. Inventory portrait
For the current Query, probe quickly retrieves a set of related items
from the current inventory to inspect relevant categories
and fields. The inventory portrait is a JSON object with:
- probe_confidence, containing max_similarity, mean_similarity, and
  similarity_std for the probe results; and
- categories, a list whose records contain category, probe_density,
  field_fill_rates (a JSON map from field names to fill rates),
  coverage_gaps, and recent_changes.
inventory_portrait: {inventory_portrait}

3. Item evidence
The item evidence is a list containing the union of items
appearing in all rollout final results.
Each item record contains:
- item_id: the item's unique identifier;
- name: the item name;
- category: the item category; and
- relevant_attributes: a JSON object of relevant attributes.

Within the supplied item list, duplicate item_id values are
removed.
item_evidence: {item_evidence}

4. Rollout final results
The results are provided as a list. Each record contains:
- rollout_id: the fixed unique identifier of the independent run; and
- final_item_ids: the item IDs returned by the run. An empty list
  means that the rollout returned no result.

rollout_results: {rollout_results}

LABEL DEFINITIONS
- success: The final results support all of the user's core
  requirements and contain no clear conflict with those requirements.
- failure: The final results clearly omit or violate at least one core
  requirement, or return no result.
- uncertain: The Query is ambiguous, or the supplied item and
  inventory evidence cannot support a reliable correctness judgment.

JUDGING PROCEDURE
1. Extract the user's core intent, required capability, and explicitly
   stated objects, attributes, and constraints from the Query.
2. Inspect every rollout independently. Check whether its final items
   support all core requirements and whether they contain
   an omission, violation, or key conflict. A correct result in one
   rollout cannot supply evidence missing from another rollout.
3. Assign success, failure, or uncertain according to the definitions
   above.

OUTPUT FIELDS
Return an object with core_requirements and rollout_judgments.
- core_requirements: one string for every extracted intent,
  capability, object, attribute, or constraint.
- rollout_judgments: one record for every input rollout, containing:
  - rollout_id, copied exactly from the input;
  - label, set to success, failure, or uncertain;
  - reason, explaining how a success satisfies the requirements, what
    a failure omits, violates, conflicts with, or fails to return, or
    why an uncertain result cannot be judged reliably;
  - satisfying_item_ids, containing one or more returned items
    that clearly satisfy the requirements only for success, and an
    empty list otherwise; and
  - representative_error_item_ids, containing one or two returned
    representative errors only for a failure with such items,
    and an empty list for a no-result failure or any other label.

OUTPUT SCHEMA
Return exactly one JSON object. Do not return Markdown, analysis, or
fields outside this schema:
{
  "core_requirements": [
    "<one core intent, capability, object, attribute, or constraint>"
  ],
  "rollout_judgments": [
    {
      "rollout_id": "<input rollout ID>",
      "label": "success",
      "reason": "<supporting requirement, item, and inventory evidence>",
      "satisfying_item_ids": ["<satisfying item ID>"],
      "representative_error_item_ids": []
    }
  ]
}

OUTPUT DISCIPLINE
Use only the Query, inventory portrait, item evidence, and final
rollout results. Keep success, failure, and uncertain distinct. Do not
invent user requirements, item attributes, or inventory facts.
\end{lstlisting}

\subsection{Query-Level Summary}

\begin{lstlisting}[style=prompt,caption={Anonymized English version of the Query-level summary prompt.},label={prompt:query-summary}]
TASK
You are a search-result analyst. Given one search Query, its already
extracted core requirements, its inventory portrait, and success and
failure evidence assembled across multiple runs, summarize successful
patterns, failed patterns, and the key outcome-relevant differences.

TERMINOLOGY
- retrieval stage: Given a Query and runtime inventory evidence,
  generate and execute search routes, including the search object,
  fields, keywords, and constraints, to form a candidate item set.
- selection stage: Filter the candidate item set against the
  Query's core intent and constraints to produce the final results.

INPUTS

1. Query
query_text: {query_text}

2. Core requirements
The requirements are provided as a list of strings. Each represents
one core intent, object, attribute, or constraint already extracted
from the Query.
core_requirements: {core_requirements}

3. Inventory portrait
For the current Query, probe quickly retrieves a set of related items
from the current inventory to inspect relevant categories
and fields. The inventory portrait is a JSON object with:
- probe_confidence, containing max_similarity, mean_similarity, and
  similarity_std for the probe results; and
- categories, a list whose records contain category, probe_density,
  field_fill_rates (a JSON map from field names to fill rates),
  coverage_gaps, and recent_changes.
inventory_portrait: {inventory_portrait}

4. Successful-path evidence
This JSON object contains items and successful_rollouts.
- items is a list of items satisfying the Query's core requirements.
  Each item record contains:
  - item_id: the item's unique identifier;
  - name: the item name;
  - category: the item category; and
  - relevant_attributes: a JSON object of relevant attributes.
  Within this item list, duplicate item_id values are removed.
- successful_rollouts is a list whose records contain:
  - rollout_id;
  - reason for the success judgment;
  - satisfying_item_ids from the final result; and
  - retrieval_paths that retrieved those items.
- Every retrieval_paths record contains:
  - route: the structured search route actually used, including its
    search object, fields, keywords, and constraints; and
  - retrieved_items: correctly retrieved items, each with item_id
    and its rank within that route. If several routes retrieve the
    same item, retain its route-specific rank in every route.

successful_path_evidence: {successful_path_evidence}

5. Retrieval-failure evidence
This JSON object contains items and failed_rollouts.
- items is a list of representative erroneous items from the failed
  rollouts. Each item record contains:
  - item_id: the item's unique identifier;
  - name: the item name;
  - category: the item category; and
  - relevant_attributes: a JSON object of relevant attributes.
  Within this item list, duplicate item_id values are removed.
- failed_rollouts is a list whose records contain:
  - rollout_id and reason;
  - retrieval_action actually taken;
  - missed_reference_item_ids: items established as satisfying by
    successful-path evidence but absent from this rollout's candidate
    set;
  - representative_error_item_ids from the final result; and
  - retrieval_paths actually used.
- Every retrieval_paths record contains:
  - route: the structured search route, including search object,
    fields, keywords, and constraints; and
  - representative_error_items: representative erroneous items
    retrieved by that route, each with item_id and route-specific
    rank. This is not the complete retrieval result and may be empty.

retrieval_failure_evidence: {retrieval_failure_evidence}

6. Selection-failure evidence
This JSON object contains items and failed_rollouts.
- items is a list of satisfying reference items that reached the
  selection candidate set and items in failed final outputs. Each
  item record contains:
  - item_id: the item's unique identifier;
  - name: the item name;
  - category: the item category; and
  - relevant_attributes: a JSON object of relevant attributes.
  Within this item list, duplicate item_id values are removed.
- failed_rollouts is a list whose records contain:
  - rollout_id and reason;
  - reference_item_ids: satisfying items established by
    successful evidence and present in this rollout's selection
    candidate set;
  - final_item_ids: the complete final output in display order; and
  - representative_error_item_ids from the final result.

selection_failure_evidence: {selection_failure_evidence}

SUMMARIZATION PROCEDURE
1. Preserve the supplied core requirements; do not add requirements
   absent from the Query.
2. Consolidate equivalent successful retrieval or selection methods.
   When supported, state how many successful runs exhibit a pattern.
3. Separately consolidate omissions, violations, or incorrect
   handling in Retrieval- and Selection-failure evidence. When
   supported, state how many failed runs exhibit a pattern.
4. Compare successful and failed evidence. Identify the fields,
   keywords, constraints, candidate coverage, or selection behavior
   responsible for outcome differences, and state whether each
   difference concerns retrieval, selection, or both.
5. Retain attributes, fields, routes, and ranks that explain an
   outcome. Omit repetition and details unrelated to this Query's
   success or failure.

OUTPUT FIELDS
All fields are lists of strings; use an empty list when no evidence
supports a field.
- core_requirements: the supplied intents, objects, attributes, and
  constraints, with one requirement per string.
- inventory_portrait_summary: categories, fields, coverage, or change
  signals directly relevant to the Query's success or failure.
- successful_patterns: supported retrieval or selection patterns.
- failed_patterns: supported omissions, violations, or incorrect
  handling, with the stage identified.
- key_differences: outcome-relevant differences between successful
  and failed paths.

OUTPUT SCHEMA
Return exactly one JSON object. Do not return Markdown, analysis, or
fields outside this schema:
{
  "core_requirements": ["<one supplied core requirement>"],
  "inventory_portrait_summary": ["<one relevant portrait signal>"],
  "successful_patterns": ["<one supported successful pattern>"],
  "failed_patterns": ["<one supported failure and its stage>"],
  "key_differences": ["<one outcome-relevant difference>"]
}

OUTPUT DISCIPLINE
Use only the supplied core requirements, inventory portrait, and path
evidence. Do not invent user requirements, inventory facts, success
causes, or failure causes.
\end{lstlisting}

\noindent\textbf{Implementation note.}
Each summarization call handles one Query and does not receive its \texttt{query\_id}. Calling code associates the summary with the source Query, retaining both \texttt{query\_id} and \texttt{query\_text} for the downstream Patch proposer.

\subsection{Pairwise Replay Judge}

\begin{lstlisting}[style=prompt,caption={Anonymized English version of the pairwise replay Judge prompt.},label={prompt:pairwise-judge}]
TASK
You are a search-result quality comparator. Given one search Query,
its already extracted core requirements, its inventory portrait, and
item evidence, compare the final results in Result A and Result B.
Determine which result is better, or whether the evidence cannot
reliably distinguish them.

INPUTS

1. Query
query_text: {query_text}

2. Core requirements
The requirements are provided as a list of strings. Each represents
one core intent, object, attribute, or constraint already extracted
from the Query.
core_requirements: {core_requirements}

3. Inventory portrait
For the current Query, probe quickly retrieves a set of related items
from the current inventory to inspect relevant categories
and fields. The inventory portrait is a JSON object with:
- probe_confidence, containing max_similarity, mean_similarity, and
  similarity_std for the probe results; and
- categories, a list whose records contain category, probe_density,
  field_fill_rates (a JSON map from field names to fill rates),
  coverage_gaps, and recent_changes.
inventory_portrait: {inventory_portrait}

4. Item evidence
The list contains the union of items in the final results from
Result A and Result B.
Each item record contains:
- item_id: the item's unique identifier;
- name: the item name;
- category: the item category; and
- relevant_attributes: a JSON object of relevant attributes.

Within the supplied item list, duplicate item_id values are
removed.
item_evidence: {item_evidence}

5. Results to compare
The input is a JSON object containing:
- result_a_item_ids: item IDs returned by Result A; and
- result_b_item_ids: item IDs returned by Result B.
IDs follow final display order; an empty list means no items were
returned.
comparison_results: {comparison_results}

COMPARISON CRITERIA
- a_better: Result A supports the Query's core requirements more
  completely than Result B, or has fewer omissions, violations, key
  conflicts, or matches unsupported by inventory evidence.
- b_better: Result B meets the same criterion relative to Result A.
- uncertain: The Query is ambiguous, item or inventory evidence
  is insufficient, or the two results cannot be reliably separated
  by requirement satisfaction and error risk.

Returning no items is not automatically worse. An empty result is
usually worse when inventory evidence supports the Query and the
other result contains a clear match. It may be preferable to forcing
an unsupported match when current evidence does not support one. Use
uncertain when the comparison cannot be resolved reliably.

COMPARISON PROCEDURE
1. Check both results against every core intent, object, attribute,
   and constraint.
2. Check both for omissions, violations, key conflicts, unsupported
   matches, or no result despite clear inventory support.
3. Compare overall requirement satisfaction and error risk. Do not
   prefer a result merely because it returns more items or
   occupies a fixed A/B position.
4. Select a_better or b_better only when the evidence supports a
   reliable preference; otherwise select uncertain.

OUTPUT FIELDS
- label: a_better, b_better, or uncertain.
- reason: the decisive differences in requirement satisfaction,
  clear errors, and inventory evidence. For uncertain, explain why
  the results cannot be reliably distinguished.

OUTPUT SCHEMA
Return exactly one JSON object. Do not return Markdown, analysis, or
fields outside this schema:
{
  "label": "a_better",
  "reason": "<decisive requirement, error, and inventory differences>"
}

OUTPUT DISCIPLINE
Use only the Query, core requirements, inventory portrait, item
evidence, and final results. Do not invent user requirements, item
attributes, or inventory facts.
\end{lstlisting}

\noindent\textbf{Implementation note.}
The prompt hides which result is Current or Candidate. Calling code converts its positional label to Candidate-relative \emph{win}/\emph{loss}/\emph{uncertain} and maps disagreement across original and order-swapped reviews to \emph{uncertain}.

\end{document}